\documentclass[12pt,twoside]{article}
\usepackage{fleqn,espcrc1}

\usepackage{graphicx}
\usepackage{epsfig}
\usepackage[figuresright]{rotating}

\newcommand{\AmS}{{\protect\the\textfont2
  A\kern-.1667em\lower.5ex\hbox{M}\kern-.125emS}}

\title{The Sigma Commutator from Lattice QCD}

\author{Stewart V. Wright, Derek B. Leinweber and Anthony
        W. Thomas\address{Department of Physics and Mathematical
        Physics \\
        and Special Research Centre for the Subatomic Structure of
        Matter,\\
        University of Adelaide, Adelaide 5005, Australia}%
} 
\begin{document}

\maketitle

\vspace{-6.cm}
\hfill ADP-00-24/T407
\vspace{6.cm}

\begin{abstract}
As a direct source of information on chiral symmetry breaking within
QCD, the sigma commutator is of considerable importance.
Since hadron structure is a non-perturbative problem,
numerical calculations on a space-time lattice are currently the only
rigorous approach.
With recent advances in the calculation of hadron masses within full
QCD, it is of interest to see whether the sigma commutator can be
calculated directly from the dependence of the nucleon mass on the
input quark mass.
We show that, provided the correct chiral behaviour of QCD is
respected in the extrapolation to realistic quark masses, one can
indeed obtain a fairly reliable determination of the sigma commutator
using present lattice data.
For two-flavour dynamical fermion QCD the sigma commutator 
lies between 45 and 55 MeV based on recent data from CP-PACS and UKQCD.
\end{abstract}

\section{WHAT IS THE SIGMA COMMUTATOR ?}

In the quest to understand hadron structure within QCD, small violations 
of fundamental symmetries play a vital role. The sigma commutator,
$\sigma_{N}$:
\begin{equation}
\sigma_{N}=\frac{1}{3}\left\langle N\right| \left[ Q_{i5},\left
    [ Q_{i5},\cal {H}\right] \right] \left| N\right\rangle = 
    \bar{m}\left\langle N\right| \bar{u}u+\bar{d}d \left|
    N\right\rangle =
    \bar{m}\frac{\partial M_{N}}{\partial \bar{m}}
\label{eqn:FH-Commutator} 
\end{equation}
(with $Q_{i5}$ the two-flavour ($i$=1, 2, 3) axial charge) is an 
extremely important example of such a symmetry.

\section{PREVIOUS ATTEMPTS}

$\sigma_{N}$ cannot be accessed directly by
experimental measurements.  However, one can infer from world data
a value of $45\pm 8$ MeV \cite{Sainio}.  This result has been under
some scrutiny recently due to the progress in new determinations of
the pion-nucleon scattering lengths \cite{PSI,LET} and new phase shift
analyses \cite{Arndt,Bugg}.
The full
lattice QCD calculations upon which our work is based involve only two
active flavours, the heavier third flavour is approximated by a
renormalisation of the strong coupling constant. 
As a guide, recent work suggests that
the best value of $\sigma_{N}$ may be 8 to 26 MeV larger than
the value quoted above \cite{Kneckt}.

One can notionally use QCD to directly calculate the value of
$\sigma_{N}$, but in practice the calculation has proven to be
difficult.
Early attempts \cite{CABASINO} to extract
$\sigma_N$ from the quark mass dependence of the nucleon mass 
(using Eq.(\ref{eqn:FH-Commutator})) in quenched QCD with naive
extrapolations 
produced values in the range 15 to 25 MeV.  Attention
subsequently turned to determining $\sigma_N$ by calculating the
scalar matrix element of the nucleon 
$\langle N|\bar u u + \bar d d|N\rangle$.
There it was discovered that the sea-quark loops
make a dominant contribution to $\sigma_N$ \cite{LIU,FUKUGITA}.  
These works, based on quenched QCD simulations found values in
the 40 to 60 MeV range, which are more compatible with the
experimental values quoted earlier.

On the other hand, the  most recent estimate of $\sigma_N$, 
and the only one based on a two-flavour,
dynamical-fermion lattice QCD calculation, comes from the SESAM collaboration.
They obtain a value of $18 \pm 5$ MeV \cite{SESAM}, through a direct
calculation of the scalar matrix element 
$\langle N|\bar u u + \bar d d|N\rangle $ and the quark mass $\bar{m}$.

The fact that neither $\langle N|\bar u u + \bar d d|N\rangle $, nor
$\bar m$ is renormalisation group invariant introduces a major
difficulty in calculating the sigma commutator in this approach. 
One must reconstruct 
the scale invariant result from the product of 
the scale dependent matrix element and the scale dependent 
quark masses.  The latter are extremely difficult to determine
precisely and are the chief sources of uncertainty.
Furthermore, since lattice calculations are made at quite large pion
masses, typically above 500 or 600 MeV, one needs to extrapolate, in
the pion mass down to the physical value at 140 MeV.
An important innovation adopted by
Dong {\it et al.} was to extrapolate $\langle N|\bar u u + \bar d
d|N\rangle$ using a form motivated by chiral symmetry, namely $a + b
\bar m^{\frac{1}{2}}$.  Regrettably, the value of $b$ used was
{\em not} constrained by chiral symmetry and higher order terms of the
chiral expansion were not considered.  Furthermore, since the work was
based on a quenched calculation, the chiral behaviour implicit in the
lattice results involves incorrect chiral physics \cite{Sharpe}. 

\section{THE CURRENT CALCULATION}

Our recent work \cite{Leinweber:2000sa} was motivated by the
improvements in computing power, together with the development of
improved actions \cite{ImprovedActions}, which have led to accurate
calculations of 
the mass of the nucleon within {\em full QCD} (for two flavours) as a
function of $\bar{m}$ down to $m_{\pi}\sim 500$ MeV. 
(Since $m_{\pi}^{2}$ is proportional to $\bar{m}$ over the range
studied we choose to display all results as a function of
$m_{\pi}^{2}$.) We showed that provided
that one has control over the extrapolation of this lattice data to
the physical pion mass, one can calculate $\sigma_{N}$ from 
$\sigma_{N} = m_{\pi}^{2}\partial M_{N} / \partial m_{\pi}^{2}$ (which
is equivalent to Eq.~(\ref{eqn:FH-Commutator}) where $\bar{m}$ was used) 
at $m_{\pi} = 140$ MeV. This
approach has the important advantage that one {\em only} needs to work
with renormalization group invariant quantities.

\begin{figure}[htb]
\centering{\
\begin{minipage}[t]{80mm}
\epsfig{file=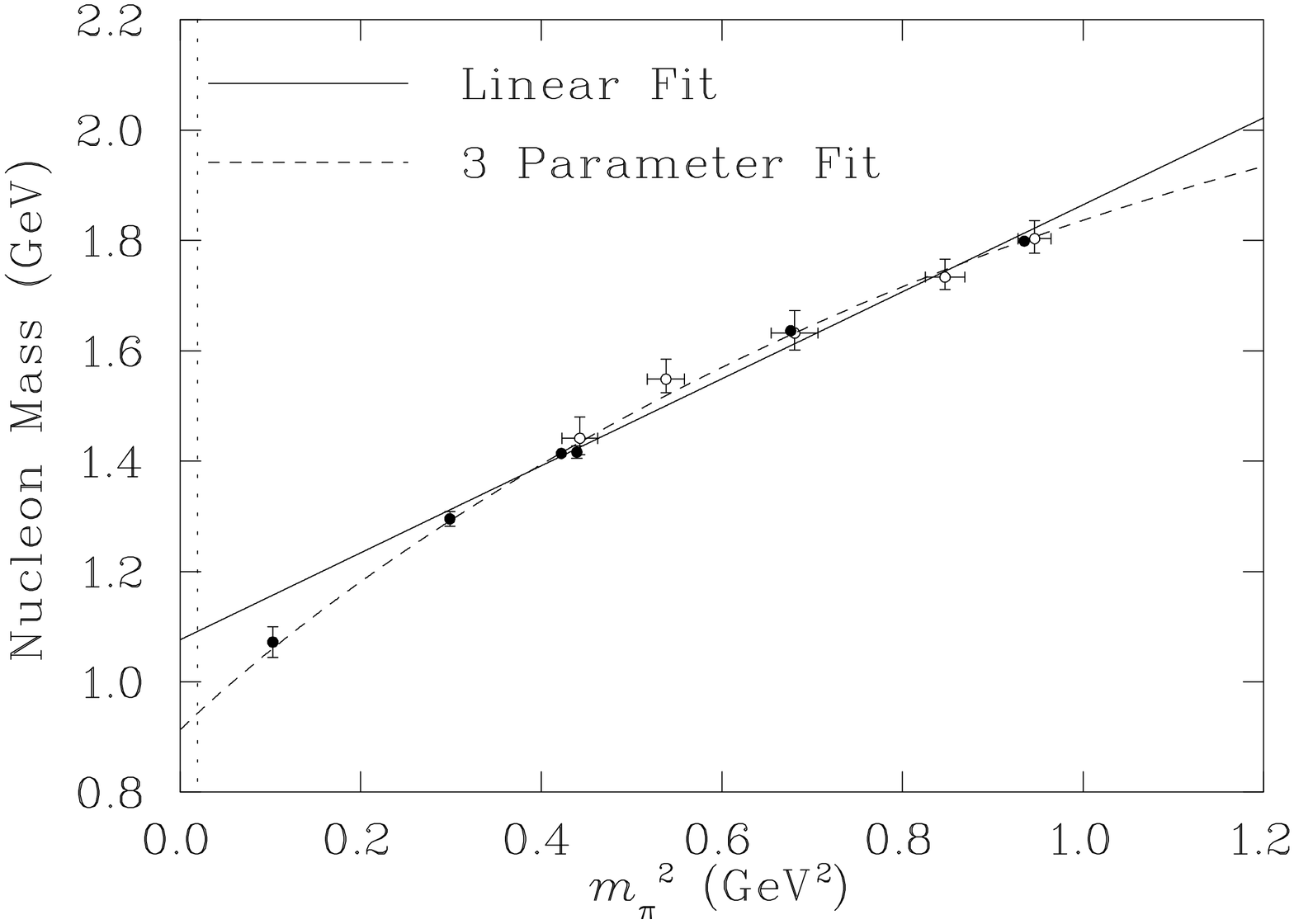,width=52mm,angle=90}
\caption{Nucleon mass versus $m_{\pi}^{2}$. 
  The solid data points are CP-PACS results
  \cite{Aoki:1999ff},
  whilst the open points are UKQCD data \cite{Allton:1998gi}. 
  Both curves are fits using Eq.~(\ref{eqn:Phen_Form_CP-PACS}).  The
  solid curve has $\tilde{\gamma} \equiv 0$, whilst the short-dashed
  curve has $\tilde{\gamma}$ unconstrained. The vertical line
  indicates the physical pion mass.
}
\label{fig:Linear Fits to Data}
\end{minipage}
\hspace{\fill}
\begin{minipage}[t]{75mm}
\epsfig{file=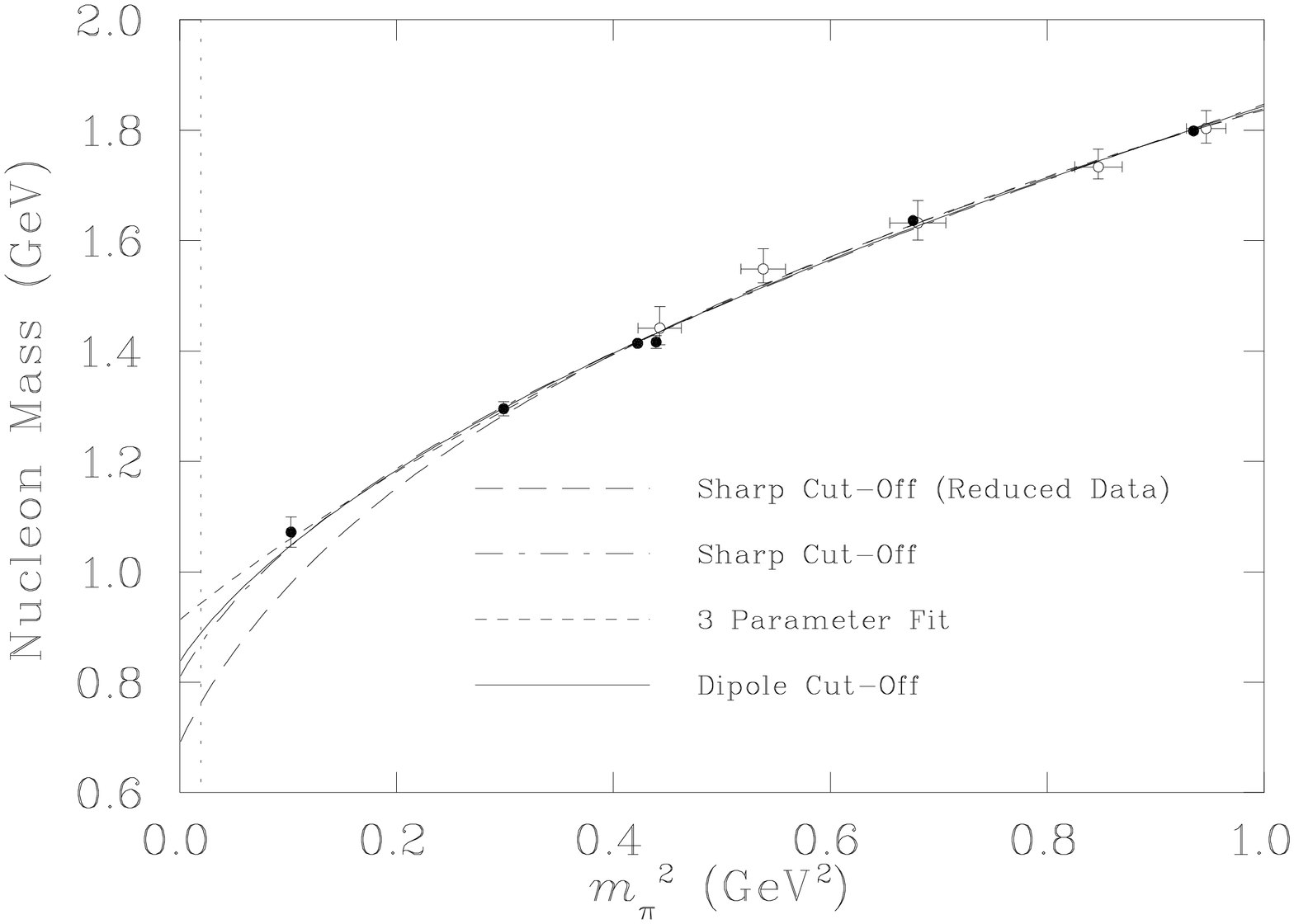,width=52mm,angle=90}
\caption{Data as labelled in Fig.~\ref{fig:Linear Fits to Data}. The
  solid curve is a fit to Eq.~(\ref{eqn:Full-Eqn}) with a dipole form
  factor, the dashed curve is the same fit using a sharp cut-off form
  factor. The long-dash curve is a fit to Eq.~(\ref{eqn:Full-Eqn})
  {\em excluding} the lowest data point.}
\label{fig:Fits to Data}
\end{minipage}}
\end{figure}

Chiral perturbation theory ($\chi$PT) predicts that the leading
non-analytic (LNA)
correction to the self energy contribution to the nucleon mass is
proportional to $m_{\pi}^{3}$ (or $\bar{m}^{3/2}$).  It can be seen in
Fig.~\ref{fig:Fits to Data} that
the preliminary point from CP-PACS \cite{Aoki:1999ff} at
$m_{\pi}^{2}\sim 0.1$ GeV$^{2}$ does indeed 
suggest some curvature in this low mass region.
These observations led the CP-PACS group to extrapolate their data with the
simple, phenomenological form:
\begin{equation}
\label{eqn:Phen_Form_CP-PACS}
M_{N}=\tilde{\alpha}+\tilde{\beta}m_{\pi}^{2}+\tilde{\gamma}m_{\pi}^{3}\, ,
\end{equation}
rather than a naive linear form ($\tilde{\gamma} \equiv 0$), as shown
in Fig.~\ref{fig:Linear Fits to Data}.
The corresponding fit to the combined data set, using
Eq.~(\ref{eqn:Phen_Form_CP-PACS}), is shown as the short-dashed curve
in Figs.~\ref{fig:Linear Fits to Data} and \ref{fig:Fits to Data}.  We
found that this fit gives 
$\sigma_{N}=29.7$ MeV. The difficulty with this purely phenomenological 
analysis was discussed in Ref.~\cite{Leinweber:1999ig}. The problem is
that a derivative is required when evaluating $\sigma_{N}$ and the
value of $\tilde{\gamma}$ found in the fit (-0.761 GeV$^{-2}$) is almost an
order of magnitude smaller than the model independent LNA coefficient, 
$\gamma ^{\mbox {\tiny LNA}}=-5.60$ GeV$^{-2}$, indicated by $\chi$PT.

\begin{figure}[b]
\centering{\
\epsfig{file=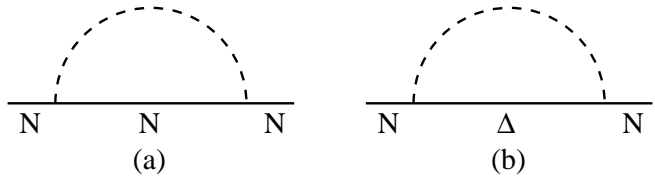}
\parbox{90mm}{
\caption{One-loop pion induced self energy of the nucleon.\label{fig:SE-N?N}}}
}
\end{figure}

Recently, an alternative approach was suggested in 
Ref.~\cite{Leinweber:1999ig}.  There it was realised that the pion
loop diagrams shown in Fig.~\ref{fig:SE-N?N} yield not only the most 
important non-analytic structure, but also give rise to the most
significant variation in the nucleon mass as $m_{\pi} \rightarrow 0$.
This leads to the following extrapolation function for $M_{N}$:
\begin{equation}
\label{eqn:Full-Eqn}
M_{N}=\alpha +\beta m_{\pi}^{2}+\sigma_{NN}(m_{\pi},\Lambda )+\sigma_{N\Delta}(m_{\pi},\Lambda )\, ,
\end{equation}
where $\sigma_{NN}$ and $\sigma_{N\Delta}$ are the
self-energy contributions of Figs.~\ref{fig:SE-N?N}(a) and
\ref{fig:SE-N?N}(b), respectively, using a cut-off in momentum
controlled by $\Lambda$.  The full analytic expressions for
$\sigma_{NN}$ and $\sigma_{N\Delta}$ are given in
Ref.~\cite{Leinweber:1999ig}. For our purposes it suffices that they
have precisely the correct LNA and next-to-leading non-analytic
behaviour required by chiral perturbation theory as $m_{\pi
}\rightarrow 0$. In addition, $\sigma_{N\Delta}$ contains the
correct, square root branch point 
($\sim [m_{\pi}^{2}-(M_{\Delta}-M_{N})^{2}]^{\frac{3}{2}}$)
at the $\Delta -N$ mass difference, which is essential for
extrapolations from above the $\Delta -N\pi$ threshold.

{}Fitting Eq.~(\ref{eqn:Full-Eqn}) to the data, including the point near 0.1
GeV$^{2}$, gives the dot-dash curve 
in Fig.~\ref{fig:Fits to Data}.
The corresponding value of $\sigma_{N}$ is 54.6 MeV 
and the physical nucleon
mass is 870 MeV. Omitting the lowest data 
point from the fit yields the long-dash
curve in Fig.~\ref{fig:Fits to Data} 
with $\sigma_{N}=65.8$ MeV, demonstrating the need for lattice
simulations of QCD at light quark masses.

\section{CONCLUSION}

The importance of the inclusion
of the correct chiral behaviour is clearly seen by the fact that it increases
the value of the sigma commutator from the 30 MeV of the unconstrained
cubic fit to around 50 MeV.  
Nevertheless, it is a remarkable result that the present lattice data for 
two-flavour dynamical-fermion QCD, yields a stable
\cite{Leinweber:2000sa} and 
accurate answer for the sigma commutator, an answer which is already within
the range of the experimental values.

This work was supported by the Australian Research Council.
\providecommand{\href}[2]{#2}
\newcommand{\wwwspires}{http://www.slac.stanford.edu/spires/find/hep/www}

\end{document}